\documentclass{moriond}

\def\be{\begin{equation}}
\def\ee{\end{equation}}
\def\bea{\begin{eqnarray}}
\def\eea{\end{eqnarray}}

\usepackage{amssymb,amsbsy,xspace,epsfig,graphicx,color,mathrsfs,amsmath,fancybox}
\usepackage{hyperref}	% This is for pdftex
\usepackage{natbib}
\usepackage{multirow}
\usepackage{rotating}

\def\Fermi{{\sc Fermi-LAT}\xspace} %{{\sc Fermi-LAT}}

\def\PAMELA{{\sc Pamela}\xspace}
\def\AMS{{\sc Ams-02}\xspace}

\def\HESS{{\sc Hess}}

\def\DAMPE{{\sc Dampe}\xspace}
\def\VERITAS{{\sc Veritas}\xspace}
\def\CALET{{\sc Calet}\xspace}
\def\MAGIC{{\sc Magic}\xspace}

\usepackage{twoopt}
\makeatletter

\newcommandtwoopt{\citeads}[3][][]{\href{http://adsabs.harvard.edu/abs/#3}
    {\def\hyper@linkstart##1##2{}
    \let\hyper@linkend\@empty\citealp[#1][#2]{#3}}}

\newcommandtwoopt{\citepads}[3][][]{\href{http://adsabs.harvard.edu/abs/#3}
    {\def\hyper@linkstart##1##2{}
    \let\hyper@linkend\@empty\citep[#1][#2]{#3}}}

\newcommandtwoopt{\citetads}[3][][]{\href{http://adsabs.harvard.edu/abs/#3}
    {\def\hyper@linkstart##1##2{}
    \let\hyper@linkend\@empty\citet[#1][#2]{#3}}}

\makeatother
\newcommand{\Photo}{}

\begin{document}
\vspace*{4cm}
\title{Theoretical interpretations of DAMPE first results: a critical review}

\author{ Y. Genolini }

\address{Service de Physique Th\'eorique, Universit\'e Libre de Bruxelles,\\ Boulevard du Triomphe, CP225, 1050 Brussels, Belgium}

\maketitle\abstracts{The \DAMPE experiment recently published its first results on the lepton ($e^+ + e^-$) cosmic-ray (CRs) flux. These results are of importance since they account for the first direct detection of the lepton \textit{break} around the energy of 1 TeV and confirm the discoveries of ground-based Cherenkov detectors. Meanwhile they reveal a new high-energy feature in the spectrum which triggered a lot of excitement on the theory side, when interpreted as the typical signature of leptophilic dark-matter annihilation. In this proceeding I mainly focus on the theoretical understanding of the lepton \textit{break}. Then I quickly review the status of the more speculative line-like DAMPE excess, whose astrophysical (pulsar) or exotic (dark matter) explanation is strongly constrained by multi-messenger astronomy.}

\section{Introduction}

This last decade, the multiplication of cosmic ray (CR) detectors brought a wealth of new data, increasing the precision of CR flux measurements in the energy range [1 GeV- 10 TeV]. The lepton flux is no exception. On the \textit{low} energy side, the experiments \PAMELA, \Fermi, and \AMS successively achieved better accuracy, reaching up to the TeV energy. On the other hand, on the \textit{high} energy side, the ground-based Cherenkov telescopes such as \HESS, \MAGIC and \VERITAS successfully probed the [500 GeV, 5 TeV] energy range, with high statistics but large systematic uncertainties. While there was already a clear hint in the first results from \HESS\citepads{2008PhRvL.101z1104A} for a steepening in the lepton spectrum, no direct measurement of this feature had never been reported so far. 

Thanks to its large acceptance (0.3 m$^2$.sr at 10 GeV) and its deep calorimeter (32 radiation lengths), \DAMPE is the first experiment which bridges the energy gap between Cherenkov telescopes and previous space missions. It is worth mentioning that the first publication of the lepton flux by \DAMPE\citepads{2017Natur.552...63D} was directly followed by the one of \CALET\citepads{2017PhRvL.119r1101A}, which is operating aboard the ISS and probing a similar energy range. The main component of \DAMPE  used for electrons identification is the BGO imaging calorimeter capable to disentangle electrons from protons thanks to the electromagnetic shower shape. This device can achieve a percent accuracy in the energy reconstruction up to 100 GeV. \\ 

Three noticeable features are shown by \DAMPE data:

\begin{itemize}
    \item A softening in the lepton flux at energies above 1 TeV.\\
    \item A slight systematic tension with \AMS and \CALET data above 50 GeV.\\
    \item A line-like signal around the energy of 1.4 TeV.\\  
\end{itemize}

The second point is particularly intriguing, since the data do not seem to follow any energy trend previously reported by other experiments. Unfortunately this discrepancy is barely addressed in a single sentence of the original paper\citepads{2017Natur.552...63D}, invoking potential uncertainties in the absolute energy scale.
In the following I comment upon the first and third points and their theoretical understanding. 

\section{Confirmation of the \textit{break} in the lepton flux}

The main claim of \DAMPE results is undoubtedly the first direct measurement of the lepton \textit{break}. The statistical significance of the evidence for broken the power-law compare to a single power-law spectrum is high and above 5 $\sigma$ applying Wilk's theorem\citepads{2017PhRvL.119r1101A}. A careful bayesian analysis\citepads{2017arXiv171205089F} leads to Bayes factor above $10^{10}$.

Theoretically, a softening in the lepton is expected in the standard picture of CR propagation. Indeed at TeV energies lepton are undergoing strong energy losses dominated by synchrotron radiations in the galactic magnetic field, and by inverse compton emission (Klein-Nishina regime) when scattering on CMB photons (see e.g.\citepads{2009A&A...501..821D}). The typical time scale $\tau_{\rm loss}$ of energy losses for high-energy leptons to reach the TeV energy is [1.2, 2.0] $10^5\;$yr. Correspondingly the diffusion length of such particle is 
$$r_{\rm diff}\sim \sqrt{4\;K(E) \;\tau_{\rm loss}}\sim [0.8, 1.2]\; \rm kpc\;,$$
considering extreme values given by the B/C ratio for the diffusion coefficient $K(E)\propto E^\delta$\citepads{2001ApJ...555..585M}. In the current paradigm, CRs are accelerated in supernovae remnants (SNRs) shocks with a rate $\nu$ of about 3 SNRs/centuary \footnote{Pulsar Wind Nebula (PWN) might also account for the lepton flux, for more extended discussion on the differences with SNRs see\citepads{2014JCAP...04..006D}}. Under the hypothesis of a homogeneous spread of the SNRs in space-time, the number of sources enclosed in the diffusion space-time volume of TeV energy positrons can be expressed as: $$N_{SNRs}\sim \frac{\nu}{{\cal V}_{\rm Gal}} \times 2\pi r_{\rm diff}^2\,h \times \tau_{\rm loss} \sim [6, 22] \left(\frac{E}{1\;{\rm TeV}}\right)^{\delta-2} \rm \;sources\,, $$ where $h$ is the half height of the Galactic disk ($\sim$ 0.1 kpc) and ${\cal V}_{\rm Gal}$ its volume ($\sim$ 250 kpc$^3$). 

Notice that the number of sources for TeV energy positrons is rather small (see also\citepads{2011ASSP...21..624B}), which means that the flux is only dominated by few sources. Regarding the energy dependence of this number, for typical values of $\delta$ ($\sim$ [0.4-0.6]), the TeV energy is a turning point above which the number of sources tends to zero in the following decade in energy. In the usual computation of CR fluxes, where sources are considered as continuously spread (jelly-like) in space time, we thus expect a softening in the lepton flux spectrum from the constraint that no sources is shining nearby. For quantitative illustrations refer for instance to Fig.1 of\citepads{1995A&A...294L..41A} where it is shown that 90\% of the flux at TeV comes from region closer than 1 kpc from the Earth. Using such a constraint, however, the shape of the flux does not exactly match the data shape within current precision. Needless to say, taking into account local sources locations and ages, as well as the injection spectra is unavoidable to interpret this feature. In particular, an exponential high-energy cut-off of the injected spectrum around 2 TeV seems to be a key ingredient is to account for the sharpness of the observed break\citepads{2014JCAP...04..006D}.

Thus, the main difficulty in interpreting the \textit{break} shape, and hopefully to increase or knowledge on CR sources, is to know  precisely which lepton sources are dominating the flux at the TeV energy. Catalogs such as the \textit{Green Catalog}\citepads{2009BASI...37...45G} or \textit{ATNF Catalog}\;\footnote{\url{http://www.atnf.csiro.au/people/pulsar/psrcat/}} have been stacking the known sources and can be used in the computation of the lepton flux. Although, even for the closest sources, their degree of completeness is questionable for the only reason that $\tau_{\rm loss}$ is about one order of magnitude above the activity timescale of a source which could be too dim to be detected (other reasons are exposed in\citepads{2011JCAP...02..031M}). One alternative is to rely on a statistical treatment, in which the sources location and age are considered as random variables\citepads{2011JCAP...02..031M, 2017A&A...600A..68G}. Obviously, fluctuations of the flux at those energies are large, and without other constraints, it seems difficult to learn any generic source property.

Recently, \Fermi published new upper bounds on the lepton dipole anisotropy\citepads{2017PhRvL.118i1103A}. Combined with radio data and the measured lepton flux around the TeV-energy, these bounds are already constraining the emission from local sources\citepads{2018arXiv180301009M}. Refinements of these bounds (and hopefully detection of anisotropy) as well as a better understanding and modelling of CR local propagation give hopes to be able to tackle sources properties and so to do astronomy thanks to the lepton flux.

\section{A peak in the lepton flux?}

A much less expected feature in the lepton flux, although much more speculative, is hinted by \DAMPE data\citepads{2017Natur.552...63D}: a line-like signal shows up at 1.4 TeV. No comments on the significance of this line can be found in the original paper, except that it should be confirmed or not by the next release. An independent careful study of the significance of this peak\citepads{2017arXiv171205089F} reports a frequentist 2.3 $\sigma$ significance (taking into account the \textit{look elsewhere} effect), and a Bayes factor of order 2. Although its significance is quite low, most of the excitement from the \DAMPE results are related to the explanation of this excess. 

The shape and the magnitude of the observed peak is actually quite constraining on its origin. Firstly, the line shape involves quasi-monochromatic injection of leptons. Secondly, as mentioned earlier, leptons are very sensitive to energy losses. This would have the effect to broaden the peak if the source were too far. Thus, in order to neglect the energy losses for 1 TeV leptons, the source must be much closer than $r_{diff}\sim 1\;$kpc defined previously (see \citepads{2017arXiv171200005H}). The source must be local and so complies with all astrophysical constraints available.

As for the explanation of the so-called \textit{positron excess}, the two preferred culprits of high-energy lepton production are DM annihilation or acceleration in pulsar environments.

\subsection{Electrophilic Dark matter?}

Counting the number of articles written on DAMPE interpetations, the annihilation of leptophilic DM is by far the most popular option. Proposed models correspond to extensions of the SM model with additional symmetries, and new charges are chosen to make DM particles annihilating exclusively into leptons. The new symmetries invoked are,  U(1) symmetry (scalar mediator:\citepads{2017arXiv171111058T,2017arXiv171111012D}, Z prime:\citepads{2017arXiv171111452C, 2017arXiv171201244C,2017arXiv171111182C,2017arXiv171111563D,2017arXiv171201239G,2017arXiv171110995F,2017arXiv171200941N}, generic vector gauge boson\citepads{2017arXiv171111579L,2017arXiv171111333G,2017arXiv171111012D}, Dark photon\citepads{2017arXiv171111000G}), $Z_2$ symmetry\citepads{2017arXiv171202021D,2017arXiv171200869L,2017arXiv171202381L,2017arXiv171200037C}, SU(2)\citepads{2017arXiv171200793C}), or extra-dimension\citepads{2017arXiv171201143Z}. {Most models try to embed the proposed DM candidates in a more general framework, also tackling some other open problems} such as neutrino masses generation (radiatively or through type II see-saw mechanism) and leptogenesis. Note that other papers were addressing the DM interpretation in a more model independent way\citepads{2017arXiv171200362J,2017arXiv171111376A,2017arXiv171111052Z,2017arXiv171200372N,2017arXiv171110989Y}. The ``electrophilic'' nature of the DM is a crucial ingredient to obtain the required peaked shape, since any sizable coupling to other SM particle (even to other lepton flavors) would generate a much broader spectrum.

Moreover, in order to avoid the peak to be smeared out, the annihilation must take place in our close environment. To get a sufficient luminosity, the DM over-density needs to be large, and one can estimate the probability of such a fluctuation. In\citepads{2017arXiv171110989Y} the authors investigate two scenarios corresponding to two extreme sizes of clump. Fixing the annihilation cross-section to the thermal one, they conclude that a clump of typical size 10 pc and 100 pc requires over-densities of 1000 and [17-35] respectively. Such large values are excluded by N-body galactic simulations (see e.g.\citepads{2009PhRvD..80c5023B} or\citepads{2010PhRvD..81d3532K}).  

Hence standard scenarios for structure formation are not likely to reproduce the require DM density to explain \DAMPE data. Although one may consider more exotic mechanisms to obtain such high densities (e.g. minispike DM clump\citepads{2005PhRvL..95a1301Z} or Ultra-compact micro halos\citepads{2012PhRvD..85l5027B,2017arXiv171201724Y}), or to enhance DM annihilation rate (e.g. Sommerfeld enhacement\citepads{2004PhRvL..92c1303H}), the production of mono-energetic e$^+$e$^-$ pairs is associated with intrinsic radiations (internal bremstralung\citepads{2017arXiv171110989Y}), but also secondary particles (i.e. inverse compton gamma-rays and radio emmision from synchrotron)\citepads{2017arXiv171200005H}. The latter are subject to very stringent constraints from gamma-rays telescope and radio ones that must not be forgotten.

\subsection{`Naked' pulsars?}

The astrophysical alternative for DM involves rapidly rotating neutron stars, so-called pulsars, which (at least in their surrounding nebula) can efficiently accelerate leptons to TeV-energies. Numerous local pulsars are observed and could potentially explain \DAMPE signal (see for example Table 4 of\citepads{2010A&A...524A..51D}). Contrary to the explanation of the well-known \textit{positron excess} which can be accounted for by a broad spectral emission, the peaked shape of \DAMPE signal requires some astrophysical \textit{fine-tuning}. Indeed the mono-energetic emission needs the pulsars wind to be cold and not affected by the shock of the associated SNR\citepads{1984ApJ...283..694K,2000MNRAS.313..504B}. Assuming this scenario, in\citepads{2017arXiv171110989Y} the authors show that Geminga or Vela are plausible candidates, and that the corresponding gamma emissions seem to evade \Fermi anisotropy constraints\citepads{2017PhRvL.118i1103A, 2017arXiv171200005H}.

\section{Conclusion}
\DAMPE, the first Chinese space probe devoted to high-energy astrophyisics, has provided a confirmation that the break at 1 TeV is an intrinsic property of CR lepton flux, rather than due to any atmospheric effect or detection systematics. In the standard CR picture, this \textit{break} is most likely a source effect and probably reflect the injection properties of one or few nearby sources. The precise measurement of the lepton flux and anisotropy, and more crucially other wavelength constraints are essential to pinpoint these prominent source(s) and hopefully help to extract generic behavior of CRs production.   
Concerning the tentative line-like signal at 1.4 TeV, one has to keep in mind that its significance is low, and that such an excess is not reported by \CALET\citepads{2017PhRvL.119r1101A}. If it were confirmed, the shape of the signal is very restrictive on the usual ``DM versus pulsar'' hypotheses invoked at the time of \PAMELA \textit{positron execess}. As for that controversy, a multimessenger approach is the most promising way to test these hypotheses viability. Of course, among the ingredients involved in the puzzle solution, the community is now eagerly waiting for \DAMPE next data releases as well.

\section*{Acknowledgments}
I thank the organizers for setting up a bright  EW Moriond week, and P. Serpico for useful discussions and advices. 
\bibliography{moriond}

\end{document}